\documentclass[prd,aps,floatfix,nofootinbib,11 pt]{revtex4}
\usepackage{amsmath,graphicx,color,epsfig}

\input{tcilatex}

\begin{document}

\title{Three-player quantum Kolkata restaurant problem under decoherence}
\author{M. Ramzan\thanks{%
mramzan@phys.qau.edu.pk}}

\address{Department of Physics Quaid-i-Azam University \\
Islamabad 45320, Pakistan}
\date{\today }

\begin{abstract}
Effect of quantum decoherence in a three-player quantum Kolkata restaurant
problem is investigated using tripartite entangled qutrit states. Amplitude
damping, depolarizing, phase damping, trit-phase flip and phase flip
channels are considered to analyze the behaviour of players payoffs. It is
seen that Alice's payoff is heavily influenced by the amplitude damping
channel as compared to the depolarizing and flipping channels. However, for
higher level of decoherence, Alice's payoff is strongly affected by
depolarizing noise. Whereas the behaviour of phase damping channel is
symmetrical around 50 \% decoherence. It is also seen that for maximum
decoherence $(p=1),$ the influence of amplitude damping channel dominates
over depolarizing and flipping channels. Whereas, phase damping channel has
no effect on the Alice's payoff. Therefore, the problem becomes noiseless
one at maximum decoherence in case of phase damping channel. Furthermore,
the Nash equilibrium of the problem does not change under decoherence.%
\newline
\end{abstract}

\maketitle
\date{\today}

\address{Department of Physics Quaid-i-Azam University \\
Islamabad 45320, Pakistan}

\address{Department of Physics Quaid-i-Azam University \\
Islamabad 45320, Pakistan}

\address{Department of Physics Quaid-i-Azam University \\
Islamabad 45320, Pakistan}


\vspace*{1.0cm}

\vspace*{1.0cm}

%\date{\today}

%\newpage

\section{Introduction}

Game theory a branch of applied mathematics, was initially developed for use
in economics by von Neumann and Morgenstern [1] and important contributions
were given by John Nash [2]. It attempts to capture mathematical behavior in
strategic situations, in which an individual's success of making a choice
depends on the choice of the other players. It is usually used to model the
behavior of biological, economical and computer systems. During last few
years, a number of classical games has been converted into the realm of
quantum mechanics [3-18]. Recently, Miszczak et al. [19] has studied a qubit
flip game on a Heisenberg spin chain. They have shown that being well aware
of the dimensionality of the system, a player can achieve a mean payoff
equal to almost 1. More recently, Sharif et al. [20] has proposed the
quantum solution to a three-player Kolkata Restaurant problem. The Kolkata
Paise Restaurant (KPR) [21] is a repeated game similar to the Minority
games, played between a large number of agents having no interaction among
themselves. Since quantum minority games [22-25] has attracted much
attention in recent years. They have also been analyzed under the influence
of decoherence by Flitney and Hollenberg [26]. It is therefore, important to
study the behaviour of quantum restaurant problem in the presence of
environmental influences.

Since, it is\ not possible to completely isolate a quantum system from its
environment. Therefore, it is important to study the system-environment
dynamics in the presence of environmental effects. Quantum games may provide
a feasible platform for implementing quantum information processing in
physical systems [27] and can be used to probe the influence of decoherence
in such systems [6, 28-31]. In this connection, quantum channels provide a
natural theoretical framework for the study of decoherence in noisy quantum
communication systems. Quantum error correction [32-33] and entanglement
purifications [34] can be employed to avoid the problem of decoherence.

In this paper, the effect of quantum decoherence in a three-player quantum
Kolkata restaurant problem is studied using entangled qutrit states. By
considering different noisy qutrit channels parameterized by decoherence
parameter $p$ such that $p\in \lbrack 0,1]$. The lower and upper limits of
decoherence parameter represent the fully coherent and fully decohered
system, respectively. It is seen that for lower level of decoherence,
amplitude damping channel heavily influences the payoffs as compared to the
depolarizing and flipping channels. However, for higher level of
decoherence, the payoff is strongly affected by depolarizing noise. On the
other hand, the behaviour of phase damping channel is symmetrical around 50
\% decoherence. Furthermore, the Nash equilibrium of the problem does not
change under decoherence.

\section{Decoherence and quantum Kolkata restaurant problem}

In the Kolkata Paise Restaurant (KPR) problem, $N$ non-communicating agents
have to choose between $n$ choices. The agents receive a gain in their
utility if their choice is not too crowded, i.e. the number of agents that
made the same choice is under some threshold limit. The choices can also
have different values of utility associated with them, accounting for a
preference profile over the set of choices. Therefore, in KPR, $N$
prospective customers choose from $N$ restaurants each evening in a parallel
decision mode. Each restaurant have identical price but different rank $k$
(agreed by the all the $N$ agents) and can serve only one customer. If more
than one agents arrive at any restaurant on any evening, one of them is
randomly chosen and is served and the rest do not get dinner that evening.

For the sake of simplicity, let the three agents, Alice, Bob and Charlie
have three possible choices: security $0$, security $1$ and security $2$.
They receive a payoff $\$=1$ if their choice is unique, otherwise they
receive $\$=0$. Therefore, the game is a one shoot game, that is, it is a
non-iterative, and the agents have no information from previous rounds.
Since the agents cannot communicate, therefore, there is nothing left to do
other than randomizing between the choices. Randomization gives the agent $i$
an expected payoff of $E^{c}(\$)=4/9$, where the superscript $c$ represents
the classical strategy.

In this problem, let Alice, Bob and Charlie share a general tripartite
entangled qutrit state of the form%
\begin{equation}
\rho _{in}=f|\Psi _{in}\rangle \left\langle \Psi _{in}\right\vert +\frac{%
(1-f)}{27}I_{27}
\end{equation}%
where the parameter $f$ controls the degree of entanglement and
\begin{equation}
|\Psi _{in}\rangle =\frac{1}{\sqrt{3}}(|000\rangle +|111\rangle +|222\rangle
)
\end{equation}%
In order to analyze the effect of entanglement, another general initial
state is also considered as given below%
\begin{equation}
|\Psi _{in}\rangle =\sin \theta \cos \phi |000\rangle +\sin \theta \sin \phi
|111\rangle +\cos \theta |222\rangle  \label{init}
\end{equation}%
where $0\leq \theta \leq \pi $ and $0\leq \theta \leq 2\pi .$\ If we set $%
\theta =\pi /4,$ $3\pi /4$ and $\phi =\pm \cos ^{-1}(1/\sqrt{3})$ in the
above equation, the three-qutrit state becomes the maximally entangled
state. The strategies of the players can be defined by the unitary operator $%
U$ acting on the initial qutrit state of the problem given as [35]
\begin{equation}
U=\left[
\begin{array}{ccc}
z_{1} & \bar{\omega}_{1} & \bar{z}_{2}\omega _{3}-\bar{z}_{3}\omega _{2} \\
z_{2} & \bar{\omega}_{2} & \bar{z}_{3}\omega _{1}-\bar{z}_{1}\omega _{3} \\
z_{3} & \bar{\omega}_{3} & \bar{z}_{1}\omega _{2}-\bar{z}_{2}\omega _{1}%
\end{array}%
\right]
\end{equation}%
where
\begin{equation}
\overrightarrow{z}=\left[
\begin{array}{c}
\sin \theta \cos \phi e^{i\alpha _{1}} \\
\sin \theta \sin \phi e^{i\alpha _{2}} \\
\cos \theta e^{i\alpha _{3}}%
\end{array}%
\right]
\end{equation}%
and
\begin{equation}
\overrightarrow{\omega }=\left[
\begin{array}{c}
\cos \chi \cos \theta \cos \phi e^{i(\beta _{1}-\alpha _{1})}+\sin \chi \sin
\phi e^{i(\beta _{2}-\alpha _{1})} \\
\cos \chi \cos \theta \sin \phi e^{i(\beta _{1}-\alpha _{2})}-\sin \chi \cos
\phi e^{i(\beta _{2}-\alpha _{2})} \\
-\cos \chi \sin \theta e^{i(\beta _{1}-\alpha _{3})}%
\end{array}%
\right]
\end{equation}
where $0\leq \chi \leq \pi /2$ and $0\leq \beta _{1},\beta _{2}\leq 2\pi .$
After the action of players unitary operators the state of the game
transform to
\begin{equation}
\rho _{\acute{f}}=(U_{A}^{\dag }\otimes U_{B}^{\dag }\otimes U_{C}^{\dag
})(\left\vert \Psi _{in}\right\rangle \left\langle \Psi _{in}\right\vert
)(U_{A}\otimes U_{B}\otimes U_{C})
\end{equation}%
The evolution of the state of a quantum system in a noisy environment can be
described by the super-operator $\Phi $ in the Kraus operator representation
as [1]

\begin{equation}
\tilde{\rho}_{f}=\Phi \rho _{f}=\sum_{k}E_{k}\rho _{f}E_{k}^{\dag }
\label{E5}
\end{equation}%
where the Kraus operators $E_{i}$ satisfy the following completeness relation

\begin{equation}
\sum_{k}E_{k}^{\dag }E_{k}=I  \label{5}
\end{equation}%
We have constructed the Kraus operators for the game from the single qutrit
Kraus operators (as given in equations (9-11) below) by taking their tensor
product over all $n^{2}$ combination of $\pi \left( i\right) $ indices

\begin{equation}
E_{k}=\underset{\pi }{\otimes }e_{\pi \left( i\right) }  \label{6}
\end{equation}%
where $n$ is the number of Kraus operators for a single qutrit channel. The
single qutrit Kraus operators for the amplitude damping channel are given by
[36]

\begin{equation}
E_{0}=\left(
\begin{array}{ccc}
1 & 0 & 0 \\
0 & \sqrt{1-p} & 0 \\
0 & 0 & \sqrt{1-p}%
\end{array}%
\right) ,\ \ E_{1}=\left(
\begin{array}{ccc}
0 & \sqrt{p} & 0 \\
0 & 0 & 0 \\
0 & 0 & 0%
\end{array}%
\right) ,\ \ E_{2}=\left(
\begin{array}{ccc}
0 & 0 & \sqrt{p} \\
0 & 0 & 0 \\
0 & 0 & 0%
\end{array}%
\right)  \label{E7}
\end{equation}%
Similarly, the single qutrit Kraus operators for the phase damping channel
are given as [36]

\begin{equation}
E_{0}=\sqrt{1-p}\left(
\begin{array}{ccc}
1 & 0 & 0 \\
0 & 1 & 0 \\
0 & 0 & 1%
\end{array}%
\right) ,\ \ E_{1}=\sqrt{p}\left(
\begin{array}{ccc}
1 & 0 & 0 \\
0 & \omega & 0 \\
0 & 0 & \omega ^{2}%
\end{array}%
\right) ,  \label{7}
\end{equation}%
where $\omega =e^{\frac{2\pi i}{3}}.$ The single qutrit Kraus operators for
the depolarizing channel are given by [37]

\begin{equation}
E_{0}=\sqrt{1-p}I,\ E_{1}=\sqrt{\frac{p}{8}}Y,\ E_{2}=\sqrt{\frac{p}{8}}Z,\
E_{3}=\sqrt{\frac{p}{8}}Y^{2},\ E_{4}=\sqrt{\frac{p}{8}}YZ
\end{equation}

\begin{equation}
E_{5}=\sqrt{\frac{p}{8}}Y^{2}Z,\ E_{6}=\sqrt{\frac{p}{8}}YZ^{2},\ \ E_{7}=%
\sqrt{\frac{p}{8}}Y^{2}Z^{2},\ \ E_{8}=\sqrt{\frac{p}{8}}Z^{2}  \label{E8}
\end{equation}%
where

\begin{equation}
Y=\left(
\begin{array}{ccc}
0 & 1 & 0 \\
0 & 0 & 1 \\
1 & 0 & 0%
\end{array}%
\right) ,\ \ Z=\left(
\begin{array}{ccc}
1 & 0 & 0 \\
0 & \omega & 0 \\
0 & 0 & \omega ^{2}%
\end{array}%
\right)  \label{9}
\end{equation}%
The single qutrit Kraus operators for the phase flip channel are given by

\begin{equation}
E_{0}=\left(
\begin{array}{ccc}
1 & 0 & 0 \\
0 & \sqrt{1-p} & 0 \\
0 & 0 & \sqrt{1-p}%
\end{array}%
\right) ,\ \ E_{1}=\left(
\begin{array}{ccc}
0 & \sqrt{p} & 0 \\
0 & 0 & 0 \\
0 & 0 & 0%
\end{array}%
\right) ,\ \ E_{2}=\left(
\begin{array}{ccc}
0 & 0 & \sqrt{p} \\
0 & 0 & 0 \\
0 & 0 & 0%
\end{array}%
\right)
\end{equation}%
and the single qutrit Kraus operators for the trit-phase flip channel are
given by

\begin{eqnarray}
E_{0} &=&\sqrt{1-\frac{2p}{3}}\left(
\begin{array}{ccc}
1 & 0 & 0 \\
0 & 1 & 0 \\
0 & 0 & 1%
\end{array}%
\right) ,\ \ E_{1}=\sqrt{\frac{p}{3}}\left(
\begin{array}{ccc}
0 & 0 & e^{\frac{2\pi i}{3}} \\
1 & 0 & 0 \\
0 & e^{\frac{-2\pi i}{3}} & 0%
\end{array}%
\right) ,  \notag \\
E_{2} &=&\sqrt{\frac{p}{3}}\left(
\begin{array}{ccc}
0 & e^{\frac{-2\pi i}{3}} & 0 \\
0 & 0 & e^{\frac{2\pi i}{3}} \\
1 & 0 & 0%
\end{array}%
\right) ,\ \ E_{3}=\sqrt{\frac{p}{3}}\left(
\begin{array}{ccc}
0 & e^{\frac{2\pi i}{3}} & 0 \\
0 & 0 & e^{\frac{-2\pi i}{3}} \\
1 & 0 & 0%
\end{array}%
\right)
\end{eqnarray}%
where $p=1-e^{-\Gamma t}$ represents the quantum noise parameter usually
termed as decoherence parameter. Here the bounds $[0,1]$ of $p$ correspond
to $t=0$, $\infty $ respectively. The final state of the game after the
action of the channel can be written as
\begin{equation}
\rho _{f}=\Phi _{p}(\rho _{f})
\end{equation}%
where $\Phi _{\alpha }$ is the super-operator realizing the quantum channel
parametrized by the real number $p$ (decoherence parameter). The payoff
operator for $i^{\text{th}}$ player (say Alice) can be written as
\begin{eqnarray}
P_{A} &=&\sum\limits_{x_{3,}x_{2,}x_{1}=0}^{2}|x_{3}x_{2}x_{1}\rangle
\left\langle x_{3}x_{2}x_{1}\right\vert ,x_{3}\neq x_{2}\neq x_{1}  \notag \\
+\sum\limits_{x_{3,}x_{2,}x_{1}=0}^{2}|x_{3}x_{2}x_{1}\rangle \left\langle
x_{3}x_{2}x_{1}\right\vert ,x_{3} &=&x_{2}\neq x_{1}  \notag \\
&&
\end{eqnarray}%
The payoff of $i^{\text{th}}$ player can be calculated as
\begin{equation}
E_{i}(\$)=\text{Tr}\{P_{A}\tilde{\rho}_{f}\}
\end{equation}%
where Tr represents the trace of the matrix. The optimal strategy for
players is found to be $U_{\text{opt}}$ given by%
\begin{eqnarray}
U_{\text{opt}}(\theta ,\phi ,\chi ,\alpha _{1},\alpha _{2},\alpha _{3},\beta
_{1},\beta _{2}) &=&(\frac{\pi }{4},\cos ^{-1}(1/\sqrt{3}),\frac{\pi }{4},%
\frac{5\pi }{18},\frac{5\pi }{18},\frac{5\pi }{18},\frac{\pi }{3},\frac{%
11\pi }{6})  \notag \\
&&
\end{eqnarray}%
It is seen that the results are consistent with Ref. [20] for $p=0$.

In order to interpret the effect of decoherence on the three-player quantum
Kolkata restaurant problem, different graphs has been plotted as a function
of decoherence parameter. In figure (1), Alice's payoff is plotted as a
function of decoherence parameter $p$ for (a) $f=0.2,$ (b) $f=0.5,$ (c) $f=1$
and $\theta =\frac{\pi }{4},$ $\phi =\cos ^{-1}(1/\sqrt{3})$ for amplitude
damping, depolarizing, phase damping, trit-phase flip and phase flip
channels, where AD, Dep, PD, TPF and PF represent the amplitude damping,
depolarizing, phase damping, trit-phase flip and phase flip channels
respectively. It is seen that Alice's payoff is heavily influenced by the
amplitude damping channel as compared to the depolarizing and flipping
channels. In figures (2 and 3), Alice's payoff is plotted as a function of $%
\theta $ and $\phi $ for $p=0.3$ and $p=0.7$ (a) amplitude damping, (b)
phase damping, (c) depolarizing and (d) trit-phase flip channels,
respectively, for the state of equation (3). It is seen that for higher
level of decoherence (see figure 3), Alice's payoff is strongly affected by
depolarizing noise. Whereas the behaviour of phase damping channel remains
symmetrical around 50 \% decoherence. In figures (4), Alice's payoff is
plotted as a function of $\theta $ and $\phi $ for $p=1$ (a) amplitude
damping, (b) phase damping, (c) depolarizing and (d) trit-phase flip
channels, respectively. It is shown that for maximum decoherence i.e. $p=1,$
amplitude damping channel dominates over the depolarizing and flipping
channels having considerable reduction in the payoff. Whereas, phase damping
channel has no effect on the Alice's payoff. In case of phase damping
channel, the problem becomes noiseless at maximum decoherence. However,
maximal entanglement gives the maximum payoff\ ($6/9$ at $p=0$) and it
reduces as one changes the degree of entanglement from its maxima or
introduces the value of decoherence parameter $p>1$. Furthermore, the Nash
equilibrium of the problem does not change under decoherence.

\section{Conclusions}

Quantum three-player Kolkata restaurant problem is investigated in the
presence of decoherence using tripartite entangled qutrit states using
amplitude damping, depolarizing, phase damping, trit-phase flip and phase
flip channels. It is seen that for lower level of decoherence, amplitude
damping channel heavily influences the payoffs as compared to the
depolarizing and flipping channels. However, for higher level of
decoherence, the payoff is strongly affected by depolarizing noise. It is
also seen that for maximum level of decoherence, amplitude damping channel
dominates over the depolarizing and flipping channels. Whereas, phase
damping channel has no effect on the Alice's payoff at $p=1$. Therefore, the
problem becomes noiseless at maximum decoherence for phase damping channel
only. Furthermore, the Nash equilibrium of the problem does not change under
decoherence.

{\huge Figures captions}\newline
\textbf{Figure 1}. (Color online). Alice's payoff is plotted as a function
of decoherence parameter $p$ for (a) $f=0.2,$ (b) $f=0.5,$ (c) $f=1$ and $%
\theta =\frac{\pi }{4},$ $\phi =\cos ^{-1}(1/\sqrt{3})$ for amplitude
damping, depolarizing, phase damping, trit-phase flip and phase flip
channels.\newline
\textbf{Figure 2}. (Color online). Alice's payoff is plotted as a function
of $\theta $ and $\phi $ for $p=0.3$ (a) amplitude damping, (b) phase
damping, (c) depolarizing and (d) trit-phase flip channels for the state of
equation.\newline
\textbf{Figure 3}. (Color online). Alice's payoff is plotted as a function
of $\theta $ and $\phi $ for $p=0.7$ (a) amplitude damping, (b) phase
damping, (c) depolarizing and (d) trit-phase flip channels for the state of
equation.\newline
\textbf{Figure 4}. (Color online). Alice's payoff is plotted as a function
of $\theta $ and $\phi $ for $p=1$ (a) amplitude damping, (b) phase damping,
(c) depolarizing and (d) trit-phase flip channels for the state of equation.%
\newline
\newpage

\begin{figure}[tbp]
\begin{center}
\vspace{-2cm} \includegraphics[scale=0.6]{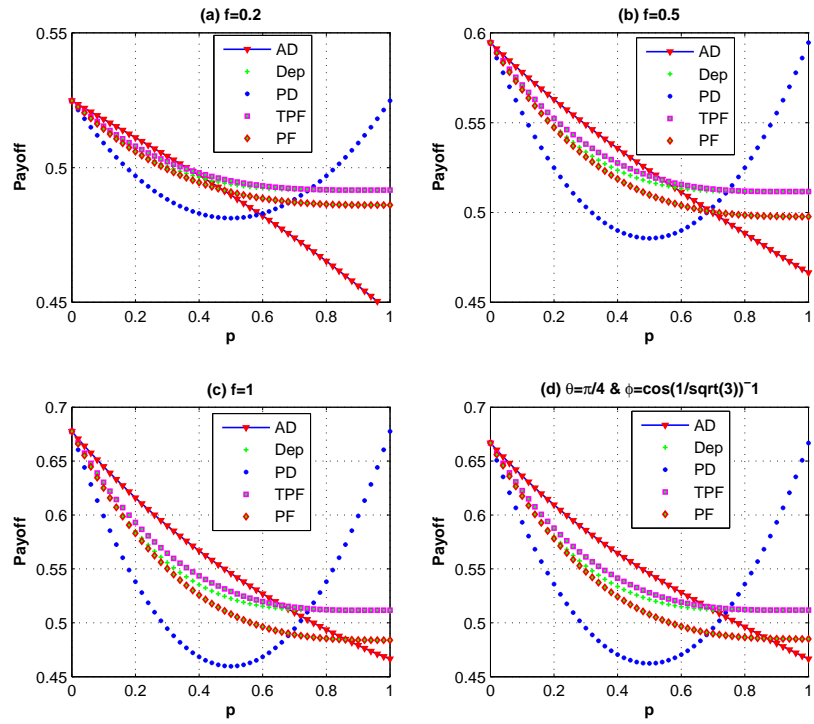} \\[0pt]
\end{center}
\caption{(Color online). Alice's payoff is plotted as a function of
decoherence parameter $p$ for (a) $f=0.2,$ (b) $f=0.5,$ (c) $f=1$ and $%
\protect\theta =\frac{\protect\pi }{4},$ $\protect\phi =\cos ^{-1}(1/\protect%
\sqrt{3})$ for amplitude damping, depolarizing, phase damping, trit-phase
flip and phase flip channels.}
\end{figure}
\begin{figure}[tbp]
\begin{center}
\vspace{-2cm} \includegraphics[scale=0.6]{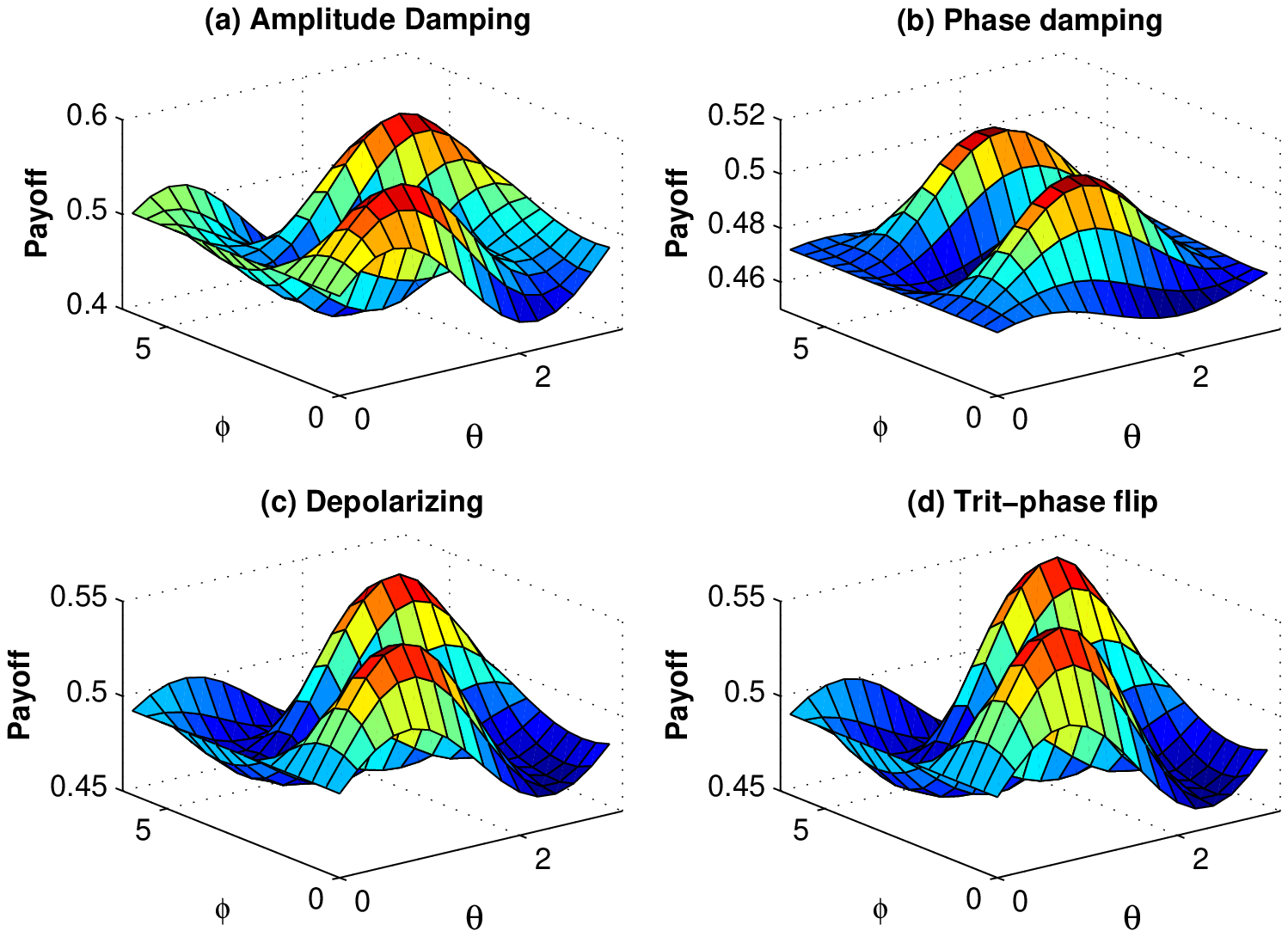} \\[0pt]
\end{center}
\caption{(Color online). Alice's payoff is plotted as a function of $\protect%
\theta $ and $\protect\phi $ for $p=0.3$ (a) amplitude damping, (b) phase
damping, (c) depolarizing and (d) trit-phase flip channels for the state of
equation.}
\end{figure}
\begin{figure}[tbp]
\begin{center}
\vspace{-2cm} \includegraphics[scale=0.6]{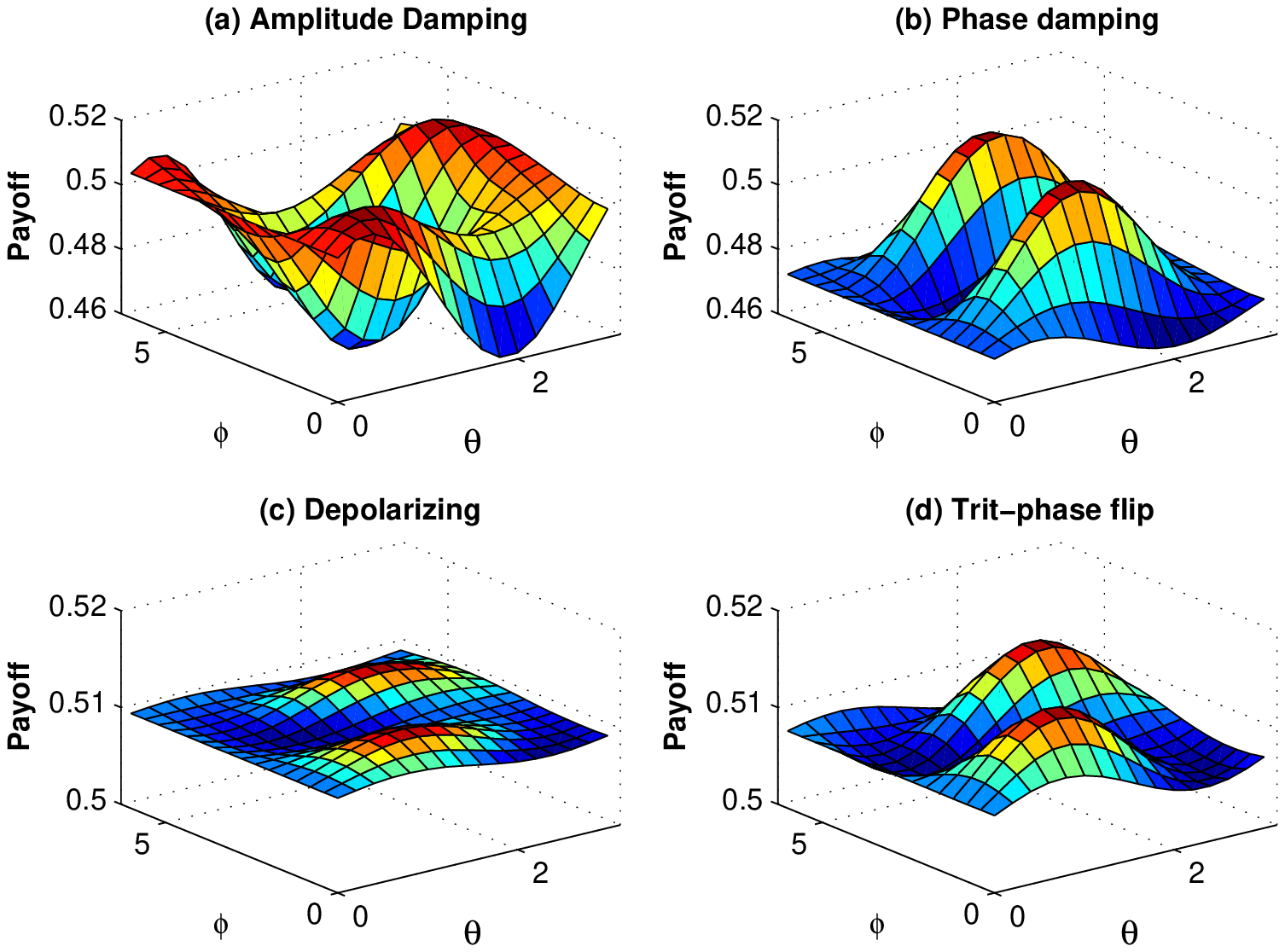} \\[0pt]
\end{center}
\caption{(Color online). Alice's payoff is plotted as a function of $\protect%
\theta $ and $\protect\phi $ for $p=0.7$ (a) amplitude damping, (b) phase
damping, (c) depolarizing and (d) trit-phase flip channels for the state of
equation.}
\end{figure}
\begin{figure}[tbp]
\begin{center}
\vspace{-2cm} \includegraphics[scale=0.6]{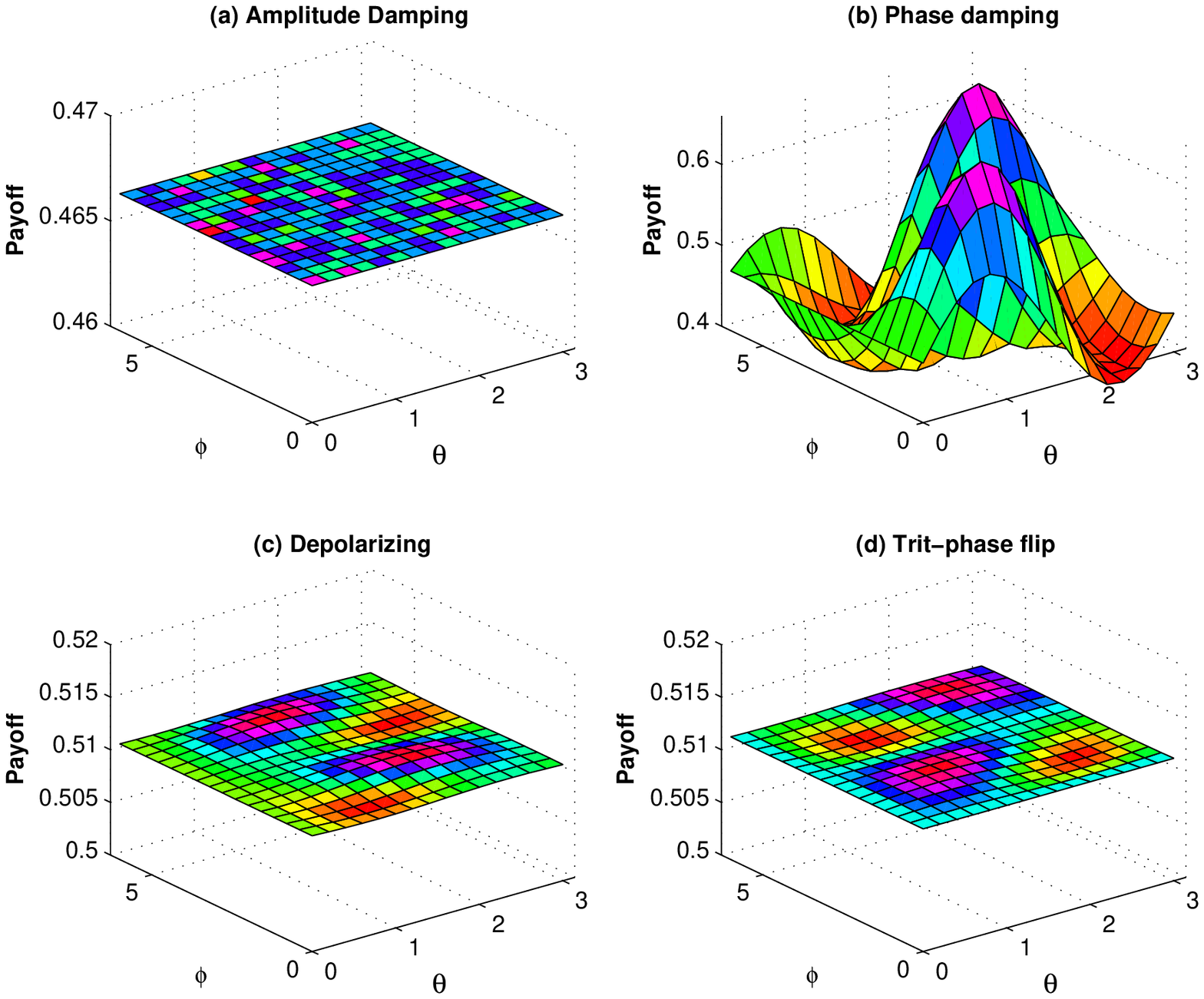} \\[0pt]
\end{center}
\caption{(Color online). Alice's payoff is plotted as a function of $\protect%
\theta $ and $\protect\phi $ for $p=1$ (a) amplitude damping, (b) phase
damping, (c) depolarizing and (d) trit-phase flip channels for the state of
equation.}
\end{figure}

\end{document}